\documentclass[prl,twocolumn,,showpacs]{revtex4}
\usepackage{graphicx}

\begin{document}

\title{Andreev-Klein reflection in graphene ferromagnet-superconductor junctions}

\author{Malek Zareyan, Hakimeh Mohammadpour, and Ali G. Moghaddam}

\affiliation{Institute for Advanced Studies in Basic Sciences,
P.O. Box 45195-1159, 45195 Zanjan, Iran}

\begin{abstract}
We show that Andreev reflection in a junction between
ferromagnetic (F) and superconducting (S) graphene regions is
fundamentally different from the common FS junctions. For a weakly
doped F graphene with an exchange field $h$ larger than its Fermi
energy $E_{\rm F}$, Andreev reflection of massless Dirac fermions
is associated with a Klein tunneling through an exchange field p-n
barrier between two spin-split conduction and valence subbands. We
find that this Andreev-Klein process results in an enhancement of
the subgap conductance of a graphene FS junction by $h$ up to the
point at which the conductance at low voltages $eV\ll \Delta$ is
greater than its value for the corresponding non-ferromagnetic
junction. We also demonstrate that the Andreev reflection can be
of retro or specular types in both convergent and divergent ways
with the reflection direction aligned, respectively, closer to and
farther from the normal to the junction as compared to the
incidence direction.
\end{abstract}

\maketitle

Transmission of low energy electrons through a normal-metal
-superconductor (NS) junction is realized via a peculiar
scattering process, known as Andreev reflection (AR)
\cite{andreev}. In AR an electron excitation with energy
$\varepsilon$ and spin direction $\sigma$ upon hitting the
NS-interface is converted into a hole excitation with the same
energy but opposite spin direction $\bar{\sigma}=-\sigma$. Under
AR the momentum change is of order $\varepsilon/v_{\rm F}$ which
is negligibly small for a degenerate N metal with large Fermi
energy $E_{\rm F}\gg\Delta$. Thus the hole velocity is almost
opposite to the velocity of the incident electron (since a hole
moves opposite to its momentum), which implies that Andreev
process is retro reflection. Andreev reflection results in a
finite conductance of a NS junction at the voltages below the
superconducting gap $\Delta$ \cite{BTK}.

\par
The fact that the Andreev reflected electron-hole belong to
different spin-subbands has an important consequence for Andreev
conductance when N metal is a ferromagnet (F). The exchange
splitting energy $h$ of F-metal induces an extra momentum change
$2h/v_{\rm F}$ of the reflected hole which diminishes the
amplitude of AR. As the result the subgap Andreev condutance of
ferromagnet-superconductor (FS) junctions decreases with
increasing $h$ and vanishes for a half-metal F with $h=E_{\rm F}$
\cite{beenakker95}. Suppression of AR at FS interface is a
manifestation of the common fact that ferromagnetism and spin
singlet superconductivity are opposing phenomena. In this letter,
however, we show that the situation differs significantly if the
FS junction is realized in graphene, the recently discovered
two-dimensional (2D) carbon atoms arranged in hexagonal lattice
\cite{geim04,geim05,kim05}. We find that in a graphene FS
junction, the exchange interaction can enhance the subgap Andreev
conductance, depending on the doping of F graphene. In particular
we show that at low voltage $eV\ll\Delta$ the conductance of a
graphene FS junction with a strong exchange field $h\gg E_{\rm F}$
is larger than its value for the corresponding NS structure. We
explain this effect in terms of Andreev-Klein reflection in which
the superconducting electron-hole conversion at FS interface is
accompanied with a pseudo-relativistic Klein transmission through
an exchange built p-n barrier between the two spin-split
conduction and valence subbands.

\par
Graphene is a zero-gap semiconductor with its conical valnce and
conduction bands touching each other at the corners of hexagonal
first Brillouin zone, known as Dirac points. The carrier type,
(electron-like (n) or hole-like (p)) and its density can be tuned
by means of electrical gate or doping of underlying substrate. Due
to the connection between this specific band structure and the
pseudo-spin aspect which characterizes the relative amplitude of
electron wave function in two trigonal sublattices of the
hexagonal structure, the charge carriers in graphene behave like
2D massless Dirac fermions with a pseudo-relativistic chiral
property \cite{wallace,slonczewski,haldane,geim05,kim05}.
Currently intriguing properties of graphene, which arise from such
a Dirac like spectrum, have been the subject of intense studies
\cite{castro-rmp,geim07,katsnelson07}.

\par
Among others, peculiarity of Andreev reflection in graphene NS
junctions was predicted by Beenakker \cite{beenakker06,
beenakker07}. S region with high carrier density can be produced
by depositing superconducting electrodes on top of a graphene
sheet \cite{heersche}. It was demonstrated that unlike the highly
doped graphene or a degenerate N metal, for undoped graphene with
$\Delta\gg E_{\rm F}$ the change in the momentum upon AR could be
of order of the momentum of the incident electron. In this limit
the dominant process is AR of an electron from the conduction band
into a hole in the valence band in which the reflection angle
(versus the normal to the NS interface) is inverted with respect
to the incidence angle, making Andreev process a specular
reflection. Transition from retro reflection at $\Delta\ll E_{\rm
F}$ to specular AR at $\Delta\gg E_{\rm F}$ is associated with an
inversion of the voltage dependence of the subgap Andreev
conductance.

\par
Recently proximity induced ferromagnetism was experimentally
realized in graphene spin-valve structures \cite{vanwees}.
Intrinsic ferromagnetic correlations were also predicted to exist
in graphene sheets \cite{peres} and nanoribbons \cite{louie}. For
a pure F graphene sheet the exchange energy shifts the normal
Fermi level at Dirac point ($E_{\rm F}=0$) upward (downward) by
$h$ in its spin-up (down) subband. A important consequence of the
gapless Dirac spectrum is that this shift makes the up and down
spin carriers to be electron-like (n type) and hole-like (p type),
respectively \cite{peres}. Concerning the transport between up and
down spin-subbands the exchange field, thus, operates as a p-n
potential barrier. The similar situation happens for a doped F
graphene ($E_{\rm F}\neq 0$) samples with large exchange energies
$h\gg E_{\rm F}$(see the right inset in Fig. \ref{fig2}). At a
graphene FS interface AR converting electron-hole from different
spin subbands will bring this exchange correlations built p-n
barrier into effect. Already reflectionless transmission of chiral
electrons through wide and high normal graphene p-n barriers was
demonstrated \cite{ando98,cheianov,katsnelson,beenakker08}. This
effect called Klein tunneling is analogous to the corresponding
effect in quantum relativistic theory \cite{klein}. We show that
the spin Klein tunneling at graphene FS junction leads to an
enhancement of the amplitude of AR and the resulting Andreev
conductance by the exchange field. This finding specific to
graphene is in striking contrast to the behaviour of Andreev
conductance of a FS junction in the ordinary metals.

\par
Concerning the connection between the incidence and reflection
directions, we further demonstrate variety of Andreev processes
taking place in graphene FS junctions. For an incident
spin-$\sigma$ electron with the velocity direction angle
$\phi_\sigma$ versus the normal to the junction, Andreev
reflection can be of retro or specular types indicated,
respectively, with or without a sign change in the reflection
angle, ${\rm sign}[\phi'_{\bar{\sigma}}]=\pm{\rm
sign}[\phi_\sigma]$, in both convergent
$|\phi'_{\bar{\sigma}}|<|\phi_\sigma|$ and divergent
$|\phi'_{\bar{\sigma}}|>|\phi_\sigma|$ ways. The type of AR
depends on the ratio $h/E_{\rm F}$, the spin $\sigma$ and energy
$\varepsilon$ of the incident electron.

\par
We consider a wide graphene FS junction normal to $x$-axis with
ferromagnetic region for $x<0$ and highly doped superconducting
region for $x>0$. In the F region the two up and down
($\sigma=\pm$) spin subbands are split by $2h$, such that the spin
$\sigma$ excitation spectrum versus 2D wave vector ${\bf
k}_{\sigma}=(k_{\sigma},q_{\sigma})$ is give by
\begin{eqnarray}
\varepsilon_{\sigma}=|E_{\rm F}\pm\hbar v |{\bf
k}_{\sigma}|+\sigma h|,
 \label{esigma}
\end{eqnarray}
where the two branches $\pm$ of the spectrum originate from the
valence and conduction bands, respectively. In S region $h=0$ and
the superconducting correlations are characterized by the order
parameter $\Delta$ which is taken to be real and constant. For
calculation we adopt Dirac-Bogoliubov-de Gennes equation
\cite{beenakker06} which describes the superconducting correlation
between massless Dirac fermions with different valley indices. In
the presence of an exchange interaction it has the form

\begin{equation}
           \left(
   \begin{array}{cc}
     H_{0}-\sigma h & \Delta \\
     \Delta^{\ast}& -(H_{0}-\bar{\sigma} h)
     \\
   \end{array}
 \right)
 \left(
           \begin{array}{c}
           u_{\sigma}\\
           v_{\bar{\sigma}}
           \end{array}
           \right)
           =\varepsilon_{\sigma}\left(
           \begin{array}{c}
           u_{\sigma}\\
           v_{\bar{\sigma}}
           \end{array}
           \right),
\label{hamiltonian}
\end{equation}
where $H_{0}=-i\hbar v_{\rm
F}(\sigma_{x}\partial_{x}+\sigma_{y}\partial_{y})-U(\vec{r})-E_{\rm
F}$ is the Dirac Hamiltonian and the potential energy $U(r)=U_0\gg
E_{\rm F}$ in S and $U(r)=0$ in F; $\sigma_x$ and $\sigma_y$ are
Pauli matrices in the pseudospin space of the sublattices.

\par
Within the scattering formalism we find the spin-dependent
amplitude of Andreev and normal reflections from FS interface. An
incident spin-$\sigma$ electron from left to FS interface with a
subgap energy $\varepsilon \leq \Delta$ and the incidence angle
$\phi_{\sigma}=\arcsin(\hbar v_{\rm F} q/(\varepsilon+E_{\rm
F}+\sigma h))$ can be either normally reflected or Andreev
reflected as a hole with opposite spin $\bar{\sigma}$ along the
reflection angle $\phi'_{\bar{\sigma}}=\arcsin(\hbar v_{\rm F}
q/(\varepsilon-E_{\rm F}+\sigma h))$. Denoting the amplitude of
normal and Andreev reflections, respectively, by $r_{\sigma}$ and
$r_{A\sigma}$ the wave function inside F is written as
$\Psi_{F\sigma}=\psi_{e\sigma}^{+}+r_{\sigma}\psi_{e\sigma}^{-}+r_{A\sigma}\psi_{h\bar{\sigma}}^{-},$
where $\psi_{e\sigma}^{\pm}\propto\exp(iq\pm
ik_{\sigma}x)\times(\exp(\mp i\phi_\sigma),\pm 1,0,0)$ and
$\psi_{h\bar{\sigma}}^{-}\propto\exp(iq-
ik'_{\bar{\sigma}}x)\times(0,0, \exp(i\phi'_{\bar{\sigma}}),1),$
are the eigenstates of Hamiltonian (\ref{hamiltonian}) in F. The
transmitted part of the electron into S,
$\Psi_{S\sigma}=a_{\sigma}\psi_{S}^{+}+b_{\sigma}\psi_{S}^{-}$
consists of two superconducting quasiparticles whose wave
functions $\psi_{S}^{\pm}=\exp(iq+ ik_{S\pm}x)\times(\exp(\pm
i\beta),\pm \exp(\pm i\beta),1,\pm 1)$ decay exponentially as a
function of $x$ ($\beta=\arccos(\varepsilon/\Delta)$). Matching
the wave functions in F and S at the interface $x = 0$ we obtain
\begin{eqnarray}
r_{\sigma}=\frac{\sec\beta\sqrt{\cos(\phi_{\sigma})\cos(\phi'_{\bar{\sigma}})}}{
\cos(\phi'_{\bar{\sigma}}-\phi_{\sigma})/2+i\tan\beta
\cos(\phi'_{\bar{\sigma}}+\phi_{\sigma})/2},\,\,\,\label{r}\\
r_{A\sigma}=\frac{-\sin(\phi'_{\bar{\sigma}}+\phi_{\sigma})/2+i\tan\beta
\sin(\phi'_{\bar{\sigma}}-\phi_{\sigma})/2}
{\cos(\phi'_{\bar{\sigma}}-\phi_{\sigma})/2+i\tan\beta
\cos(\phi'_{\bar{\sigma}}+\phi_{\sigma})/2}.\label{ra}
\end{eqnarray}
From the conservation of the $y$-component wave vector $q$ under
the scattering we obtain the following relation between the
incidence and reflection angles
\begin{equation}
\frac{\sin\phi'_{\bar{\sigma}}}{\sin\phi_{\sigma}}=\frac{\varepsilon+E_{\rm
F}+\sigma h}{\varepsilon-E_{\rm F}+\sigma h}.\label{angle}
\end{equation}
\begin{figure}
\begin{centering}\includegraphics[height=6cm]{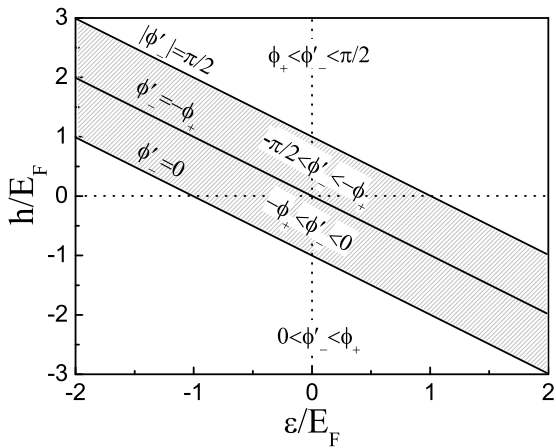}\par\end{centering}
\caption{\label{fig1}{\small Map of the Andreev reflection angle
$\phi'_{-}$ for an up-spin electron incident with the angle
$\phi_{+}$ on the grapgene FS junction. It shows dependence on the
electron energy $\varepsilon/E_{\rm F}$ and the exchange energy
$h/E_{\rm F}$ in the scale of Fermi energy. In the shaded region
between the two lines $\varepsilon/E_{\rm F}+h/E_{\rm F}=\pm1$ the
Andreev reflection is retro while in the white region it is
specular. The line $\varepsilon/E_{\rm F}+h/E_{\rm F}=0$, on which
the reflection is perfectly retro, divides the regions of
convergent reflection(below) and divergent reflection(above).
 }}
\end{figure}
\par
Inspection of the results given by Eqs. (\ref{esigma}) and
(\ref{r})-(\ref{angle}) reveals variety of spin-dependent Andreev
processes in the FS graphene junction. Consider an spin-up
electron with the energy $\varepsilon$ above the Fermi level in
the conduction band ($E_{\rm F}>0$) hitting the junction in an
angle $\phi_+$ versus the normal to the junction (negative
$x$-axis). It can be Andreev reflected as a hole with the same
energy below the Fermi level in down-spin subband. From relation
(\ref{esigma}) we see that as long as $\varepsilon+h\leq E_{\rm
F}$ the hole is in the conduction band too, and AR is retro. Upon
AR $q$ and $\varepsilon$ are conserved but the magnitude of the
momentum is changed by $2(\varepsilon+h)/v_{\rm F}$ (see Eq.
(\ref{esigma})). These conditions result in the relation
$k'_{-}=\sqrt{(\varepsilon+h-E_{\rm F})^2-(\varepsilon+h+E_{\rm
F})^2\sin^2\phi_+},$ for the $x$-component of the down spin hole
which shows that above the critical angle
$\phi^{c}_+=\arcsin(|\varepsilon-E_{\rm F}+h|/|\varepsilon+E_{\rm
F}+h|)$, the hole wave function is evanescent and amplitude of AR
vanishes. When $\varepsilon+h=0$, $\phi_{+}^{c}=\pi/2$ and the
$x$-component of the reflected hole is the same as the incident
electron, so $\phi'_-=-\phi_+$. This defines a line in the phase
space of $\varepsilon/E_{\rm F}$ and $h/E_{\rm F}$ (see Fig.
\ref{fig1}) on which AR is perfectly (incidence and reflection
directions are aligned) retro. For a fixed $\varepsilon$,
increasing $h$ results in an increase in $|\phi'_-|$, implying a
divergent retro reflection (see Eq. (\ref{angle})). As $h$
approaches $E_{\rm F} -\varepsilon$ the angle
$\phi'_-\rightarrow-\pi/2$ and simultaneously $\phi^{c}_+$
decreases monotonically to approach zero. This implies that close
to the line $\varepsilon+h=E_{\rm F}$ only normally incident
spin-up electrons have a finite AR amplitude. Thus in $h/E_{\rm
F}-\varepsilon/E_{\rm F}$ phase diagram shown in Fig. \ref{fig1}
the region between two lines $\varepsilon+h=0$ and
$\varepsilon+h=E_{\rm F}$ defines the phase of divergent retro AR.
Increasing $h$ further we cross the line $\varepsilon+h=E_{\rm F}$
which is associated with a sign change in the reflection angle
from $-\pi/2 $ to $\pi/2$ implying a transition to the regime of
specular AR. For higher $h$, $\phi'_-$ decreases toward $\phi_+$
and $\phi^{c}_+$ increases. In the limit of $h+\varepsilon\gg
E_{\rm F}$, $\phi'_-\rightarrow \phi_+$ and AR becomes perfectly
specular. This defines the region of divergent specular AR above
the line $\varepsilon+h=E_{\rm F}$.
\par
In a similar way we observe that starting from the perfectly retro
AR line and decreasing $h$ for a fixed $\varepsilon$, AR remains
retro but becomes convergent with $|\phi'_-|<|\phi_+|$. The retro
convergent region $-\phi_+<\phi'_-<0$ extends between the two
lines of $\varepsilon+h=0$ and $\varepsilon+h=-E_{\rm F}$.
Approaching the line $\varepsilon+h=-E_{\rm F}$, $\phi'_-
\rightarrow 0$. Upon crossing this line we again will have a retro
to specular transition, but this time in a continues way at
$\phi'_-=0$. Below the line of $\phi'_-=0$, is the region of
convergent specular AR with $0<\phi'_-<\phi_+$. For all convergent
reflection region $\phi^{c}_+=\pi/2$. Note that at zero energy the
retro or specular (depending on $h/E_{\rm F}$) AR will transform
from divergent to convergent or vice versa upon changing the spin
direction of the incident electron (see Fig. \ref{fig1}).

\par
We have calculated the Andreev conductance of FS junction at zero
temperature via Blonder-Tinkham-Klapwijk formula
\begin{eqnarray}
G=\sum_{\sigma}G_{\sigma}
\int_0^{\phi^{c}_{\sigma}}{d\phi_{\sigma}}\cos\phi_{\sigma}(1-|r_{\sigma}|^2+|r_{A\sigma}|^2),
\end{eqnarray}
where the spin-$\sigma$ normal-state conductance
$G_{\sigma}=(2e^2/h)N_{\sigma}(eV)$ and the density of states
$N_{\sigma}(\varepsilon)=|\varepsilon+E_{\rm F}+\sigma h|W/(\pi
\hbar v_{\rm F})$ ($W$ is width of the junction).
\begin{figure}
\begin{centering}\includegraphics[height=6cm]{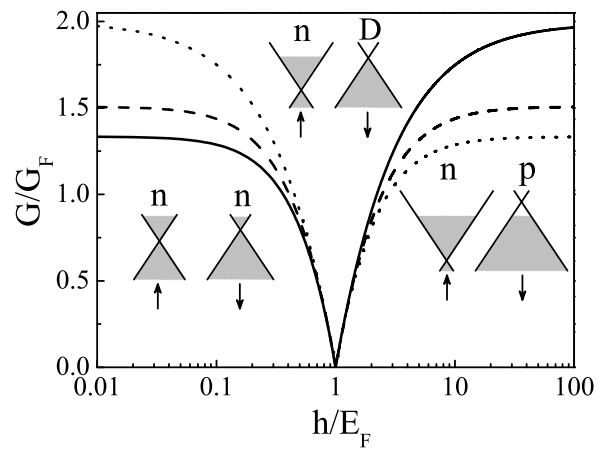}\par\end{centering}
\caption{\label{fig2}{\small Dependence of the Andreev conductance
of a graphene FS contact on the exchange field $h/E_{\rm F}$ in
units of the Fermi energy for a highly doped F, $E_{\rm
F}\gg\Delta$, and at three bias voltages $eV/\Delta=0,1/\sqrt{2},
1$ (solid, dashed and dotted curves, respectively). For $h>E_{\rm
F}$ the conductance increases with $h/E_{\rm F}$. Note that for an
undoped graphene ($E_{\rm F}=0$) the conductance for each of the
subgap voltages takes an exchange field-independent value
determined by its value for a doped sample in the limit $h/E_{\rm
F}\gg1$. The insets show the configuration (being electron-like
(n), hole-like (p) or neutral (D) at Dirac point)  of the up and
down spin-subbands for three different cases of $h/E_{\rm F}<1$,
$h/E_{\rm F}=1$ and $h/E_{\rm F}>1$.}}
\end{figure}
\par
Dependence of the resulting Andreev conductance $G/G_{\rm F}$
($G_{\rm F}=G_{+}+G_{-}$) on $h/E_{\rm F}$ is presented in Fig.
\ref{fig2} for three different bias voltages
$eV/\Delta=0,1/\sqrt{2}, 1$ and for highly doped F graphene
$E_{\rm F}/\Delta\gg1$. For $h/E_{\rm F}<1$ the conductance
decreases monotonically below the corresponding NS value $G/G_{\rm
F}(h=0)$ with $h/E_{\rm F}$ and vanishes at $h=E_{\rm F}$. In this
regime the up and down spin subbands are of the same n type (left
inset of Fig. \ref{fig2}) and the effect of exchange field is to
impose a normal barrier against AR. The resulting momentum
difference of Andreev reflected electron-hole diminishes AR
amplitude and hence the Andreev conductance.  At $h=E_{\rm F}$ the
down spin subband is at Dirac point (middle inset of Fig.
\ref{fig2}) with a vanishing density of states which results in
$G/G_{\rm F}=0$. For $h>E_{\rm F}$ Fermi level of spin down
electrons is transferred into the valence band and thus the
exchange field barrier finds a p-n characteristic with the height
$2h-E_{\rm F}$(right inset of Fig. \ref{fig2}). In this case the
conductance $G/G_{\rm F}$ increases monotonically with $h/E_{\rm
F}$. In the limit of $h/E_{\rm F}\gg1$ the exchange barrier
transforms to an almost perfect p-n barrier with the height
$\simeq 2h$ which makes possible perfect transmission of chiral
Dirac fermions between the two subbands.  The enhancing Andreev
conductance reaches a limiting maximum which depends on the bias
voltage. Importantly we see that at $eV/\Delta=0$ this limiting
value $G/G_{\rm F}(h/E_{\rm F}\gg 1)=2$ is larger than the value
for corresponding NS structure $G/G_{\rm F}(h/E_{\rm F}\ll 1)=4/3$
as shown in Fig. \ref{fig2}. This effect is in contrast to the
common view that that ferromagnetic ordering and the singlet
superconductivity are exclusive phenomena.
\par
Finally, we note that with the recent successful induction of
superconductivity \cite{heersche} and spin-polarization
\cite{vanwees} in graphene sheets, realizing graphene FS junctions
seems feasible. This will allow for the experimental observation
of the above predicted effect.
\par
In conclusion, we have demonstrated the unusual features of
Andreev reflection in graphene ferromagnet-superconductor
junctions. We have shown that depending on the ratio of the
exchange field and Fermi energy $h/E_{\rm F}$ in F graphene and
the energy and spin direction of the incident electron, the
Andreev reflection may be of retro or specular types with
possibility of being convergent as well as divergent. More
fundamentally and in contrast to the common view, we have found
that for $h>E_{\rm F}$ the exchange field enhances the Andreev
conductance of FS junction to reach a maximum value which at low
bias voltages $eV\ll \Delta$ is greater that its value for the
corresponding non-ferromagnetic structure. We have explained this
effect in terms of Andreev-Klien reflection in which the
superconducting electron-hole conversion at FS interface is
associated with a Klein tunneling through the exchange built p-n
barrier between the spin-split conduction and valence subbands.

\end{document}